\documentstyle[11pt,aaspp4,epsfig]{article}

\newcommand{\hi}{\ion{H}{1}}
\newcommand{\hii}{\ion{H}{2}}

\newcommand{\kms}{km\thinspace s$^{-1}$}
\newcommand{\bvr}{$BV\!R$}

\slugcomment{}

\lefthead{Rhode et al.}
\righthead{\hi\ Holes in Holmberg II}

\begin{document}

\title{A Test of the Standard Hypothesis for the Origin of the\\ 
\hi\ Holes~in~Holmberg~II}

\author{Katherine L. Rhode\altaffilmark{1} and 
John J. Salzer\altaffilmark{2,3}}
\affil{Astronomy Department, Wesleyan University, Middletown, CT 06459}

\author{David J. Westpfahl\altaffilmark{2}}
\affil{Department of Physics, New Mexico Institute of 
Mining \& Technology, Socorro, NM  87801}

\and

\author{Lisa A. Radice\altaffilmark{2,4}}
\affil{Astronomy Department, Wesleyan University, Middletown, CT 06459}

\altaffiltext{1}{present address: Yale University Department of
Astronomy, 260 Whitney Avenue, New Haven, CT 06520}
\altaffiltext{2}{Visiting Astronomer, Kitt Peak National Observatory,
National Optical Astronomy Observatories, which is operated by the
Association of Universities for Research in Astronomy, Inc. (AURA)
under cooperative agreement with the National Science Foundation}
\altaffiltext{3}{NSF Presidential Faculty Fellow}
\altaffiltext{4}{present address: 1960 Cate Mesa Rd., Carpinteria, CA 93013}

\begin{abstract}
The nearby irregular galaxy Holmberg II has been extensively mapped in
\hi\ using the Very Large Array (VLA), revealing intricate structure
in its interstellar gas component (\cite{puche92}).  An analysis of
these structures shows the neutral gas to contain a number of
expanding \hi\ holes.  The formation of the \hi\ holes has been
attributed to multiple supernova events occurring within wind-blown
shells around young, massive star clusters, with as many as $10-200$
supernovae required to produce many of the holes.  From the sizes and
expansion velocities of the holes, Puche et al.~assigned ages of
$\sim$10$^7$ to 10$^8$ years.  If the supernova scenario for the
formation of the \hi\ holes is correct, it implies the existence of
star clusters with a substantial population of late-B, A and F main
sequence stars at the centers of the holes.  Many of these clusters
should be detectable in deep ground-based CCD images of the galaxy.

In order to test the supernova hypothesis for the formation of the
\hi\ holes, we have obtained and analyzed deep broad-band \bvr\ and
narrow-band H$\alpha$ images of Ho$\thinspace$II.  We compare the
optical and \hi\ data and search for evidence of the expected star
clusters in and around the \hi\ holes.  We also use the \hi\ data to
constrain models of the expected remnant stellar population.  Assuming
that the \hi\ holes are created by multiple SNe, that the number of
SNe required can be determined from the observed energetics of the
holes, and that the SNe represent the high-mass population of a
cluster with a normal IMF, we show that in several of the holes the
observed upper limits for the remnant cluster brightness are {\it
strongly inconsistent with the SNe hypothesis described in Puche et
al.}  Moreover, many of the \hi\ holes are located in regions of very
low optical surface brightness which show no indication of recent star
formation.  Here we present our findings, discuss their implications,
and explore possible alternative explanations for the existence of the
\hi\ holes in Ho$\thinspace$II, including the recent suggestion that
some of the holes could be produced by Gamma-ray burst events.

\end{abstract}


\section{Introduction}

High spatial and spectral resolution observations of the neutral
hydrogen gas in nearby dwarf and Magellanic irregular galaxies have
revealed remarkably intricate and complex structures (\cite{puche92},
hereafter P92; \cite{puche94}; \cite{west94}; \cite{staveley97},
\cite{kim98}).  Numerous holes are visible in the \hi\ distributions
of these galaxies, surrounded by shells of higher density.  In
Holmberg$\thinspace$II (Ho$\thinspace$II, DDO$\thinspace$50) for
example, P92 identify 51 holes, and show that in many cases expansion
of the gas surrounding the holes is directly detectable.  Typical
expansion velocities of 4 -- 10 \kms\ have been measured, and holes as
large as 1600 pc across are present.  This hole--shell morphology for
the \hi\ gas appears to be quite common, being present in essentially
all of the nearby dwarfs studied to date with adequate spatial and
velocity resolution.

Previous studies of the \hi\ distributions in nearby spiral galaxies
such as M31 and M33 (\cite{brinks86}; \cite{deul90}) also reveal the
presence of \hi\ holes.  The origin of these features has generally
been attributed to stellar winds and supernova explosions (SNe) from
young stellar associations and clusters.  This previous work naturally
suggested that the holes in Ho$\thinspace$II and other dwarfs were
caused by a similar process.  P92 put forth the hypothesis that the
holes in Ho$\thinspace$II were due to the combined action of hot
stellar winds from O and B stars, plus the SNe shocks from the same
massive stars after they exploded at the end of their lifetimes.
Based on their measured expansion rates and hole sizes, they
determined that as much as $\sim$2 $\times$ 10$^{53}$ ergs of kinetic
energy, the equivalent of $\approx$200 SNe, was required to produce
the largest holes.  The more typical holes require a few to several
dozen SNe.

Although the SNe hypothesis for explaining the \hi\ holes in
Ho$\thinspace$II is an appealing one, it is not without problems.
Several of the \hi\ holes are located at large distances from the
center of the galaxy, in regions of both low optical surface
brightness and low \hi\ column density.  At least 18 holes are located
beyond the Holmberg radius of the galaxy (the radius where the
$B$-band surface brightness drops below 26.6 magnitudes/square
arcsec), where it is unlikely that large numbers of massive stars
could have formed. The \hi\ column densities at these radii are well
below the canonical threshold density of 10$^{21}$ cm$^{-2}$
(\cite{kennicutt89}), a further indication that star formation is
unlikely to occur at these locations in the galaxy.  Therefore, it
seemed appropriate to investigate further the SNe hypothesis of P92,
to determine whether it was possible to corroborate the general
picture or, alternatively, to rule it out.

The supernova scenario described in P92 does in fact provide us with a
direct observational test.  Those authors found that many of the
expanding holes in Ho$\thinspace$II require the kinetic energy input
of $\sim$10 to as many as 200 SNe each.  Such multiple supernova
events should only occur in massive clusters or OB associations.  Age
estimates of the \hi\ holes found in Ho$\thinspace$II have been
assigned based on their sizes and expansion velocities, and typical
ages fall in the range of 10$^7$ to 10$^8$ years.  Consequently --- if
the ages are indeed accurate --- the upper main sequence stars
(late-B, A and F) should still be present in the clusters which
produced the SN precursors.  During the brief time period since these
clusters formed, they will not have dispersed significantly, and
should still be observable as blue sources at the centers of the \hi\
holes.  Since the surface brightness level associated with the
underlying old population of stars in Ho$\thinspace$II is fairly low,
the upper main sequence population in these clusters should readily
stand out.  If these young star clusters produced, for example, 50 OB
stars which became SNe, then a Salpeter IMF would predict the
existence of at least 300 upper main sequence stars which would still
be present after 10$^8$ years.  Although this population of stars
would not be resolved from the ground (a single main sequence A0 star
would have m$_V$ = 28 and 1 arc second corresponds to 15 parsecs at
the distance of Ho$\thinspace$II), the total cluster brightness should
be m$_V$ $\sim$ 22, which would be readily detectable in deep CCD
images.

In order to look for the young clusters which would provide evidence
for the SN scenario, we decided to carry out a deep, multi-color
imaging study of Ho$\thinspace$II.  Our main goal was to obtain
accurate optical magnitudes and colors of all objects appearing in and
around the locations of the \hi\ holes found in the VLA maps.  The
information tabulated in P92 regarding the ages and energy
requirements of the holes can be used to calculate the magnitudes and
colors of the clusters that should be present in the holes if the SN
scenario is correct.  Direct comparison of these calculations and our
imaging photometry allows us to carry out a sensitive test of the SN
scenario.

In the following section, details of the optical observations carried
out for this study are discussed, along with the reduction steps
executed on the data, and a complete description of the photometric 
measurements of the \hi\ hole regions and of individual sources located 
in or around the holes.  Section~\ref{section:analysis} describes 
calculations of the sizes and brightnesses of the putative clusters made 
using the information tabulated in P92.  The last two sections of the paper 
consist of a discussion of our findings and their implications for the 
scenario proposed in P92, followed by a summary and final remarks.  

\section{Observations}
\subsection{Description and Preliminary Reductions}
\label{section:observations}

Observations of Ho$\thinspace$II were obtained in February 1994 and
April 1995 with the 0.9-meter telescope at Kitt Peak National
Observatory.  The galaxy was imaged in three broad-band filters (\bvr)
during both observing runs, and in narrow-band H$\alpha$ during the
February 1994 observing run.  The detector used was a Tektronix
2048$\times$2048 CCD (T2KA), formatted to read out only the central
1280$\times$1280 pixels.  Each pixel subtended 0.68\arcsec\ on the
sky, resulting in a total field-of-view 14.5\arcmin\ on a side.
Multiple exposures were obtained through each filter, and the
telescope was dithered between exposures, to facilitate removal of
particle events (cosmic rays) in the images.  Total integration times
were 2400 s in $B$, 1800 s in $V$, and 1200 s in $R$ for the 1994
data, and 2700 s in $B$, 1800 s in $V$, and 1800 s in $R$ for the 1995
data.  Images were taken under clear observing conditions.
Photometric standards (\cite{landolt83,landolt92}) were also observed
for use as calibration sources.  Observations of additional dwarf
galaxies were obtained during these runs (Ho I, K 73, M81dwA, IC 2574,
Leo A, Sex A, Gr8, DDO 147); the results for these objects will be
presented in a subsequent paper (\cite{slaz99}).

Images taken during the 1994 and 1995 observing runs were reduced
separately.  Preliminary reductions (overscan level subtraction, bias
image subtraction, flat field division) were carried out following
standard practices.  Multiple images taken in sequence through a
particular filter were aligned with the middle image in the sequence.
Sky subtraction was accomplished by defining a series of rectangular
regions surrounding the galaxy, measuring the mean flux level in those
regions (after masking stars and cosmic rays), and subtracting that
flux level from the image.  Multiple images in a given filter were
scaled to a common flux level (to preserve photometric integrity) and
combined into a single image, using a pixel-rejection algorithm to
eliminate cosmic rays.  The combined $B$, $V$, and $R$ images were
rotated to a north-up, east-left orientation, and their central
coordinates were determined using an astrometry routine which measures
the positions of Guide Star Catalog (\cite{lasker90}) objects
appearing in the field.

A total of six standard stars (\cite{landolt83,landolt92}) were used
to calibrate the broad-band images of Ho II from the February 1994
run, while 21 were available for the April 1995 data.  In all cases,
the photometric zero-point constants were determined with an accuracy
of $\sim$0.01 magnitude, i.e., the nights were photometric.

In order to create the deepest possible \bvr\ images with which to do
photometry of the hole regions, the images from the two runs were
convolved to a common resolution, then scaled, aligned, and combined
to create a single, deep image in each filter. The fluxes were scaled
to the April 1995 values, since the photometric calibration was deemed
to be of slightly higher quality for that run.  Photometry was carried
out on faint point sources in the resultant images to quantify the
detection limit.  The 4-$\sigma$ limit on the brightness of any point
source is $B$ $=$ 23.  The resolution (PSF FWHM) of each of the final
combined \bvr\ images is 2.4\arcsec.

A continuum-subtracted H$\alpha$ image was created using the images
taken in February 1994 through the on-band ($\lambda_o$ = 6569 \AA,
$\Delta\lambda$ = 89 \AA) and off-band ($\lambda_o$ = 6409 \AA,
$\Delta\lambda$ = 88 \AA) filters.  Both narrow-band images were
aligned with the composite \bvr\ frames and convolved to a common
resolution.  The off-band image was scaled to the on-band image by
comparing the fluxes for several bright stars in the two images, after
which it was subtracted from the on-band image to produce the final
H$\alpha$ frame.  Since the narrow-band images were acquired under
non-photometric conditions, no flux calibration was attempted.

The high-resolution (natural weight) \hi\ map of Ho$\thinspace$II from
P92 was used for this study to provide information about the hole
locations.  To facilitate comparison between the optical and radio
data, the format of the \hi\ map was modified to match that of the
optical images.  Specifically, the original \hi\ map (2\arcsec/pixel)
was rescaled to the optical image scale (0.68\arcsec/pixel).  The
section of the radio map corresponding to the area covered by the
optical frames --- i.e., having the same size and central coordinates
as the optical images --- was then extracted and used to create a new
image.  Figure~\ref{fig:hi map} shows the high-resolution \hi\ map
after it has been scaled and aligned with the optical image, and
Figure~\ref{fig:combined BVR image} shows a composite \bvr\ image,
created by combining the $B$, $V$, and $R$-band data taken in April
1995.

\subsection{Photometry}
\label{section:photometry}

\subsubsection{H$\thinspace$I Holes}

To search for evidence of star clusters at the centers of the \hi\
holes, photometric measurements were carried out on the combined $B$,
$V$, and $R$ images created by merging the data from the February 1994
and April 1995 observing runs.  These combined frames provided us with
the deepest images possible for our search.

The photometry was performed using the following strategy.  The hole
positions given in Table 5 of P92 were used as an initial guess for
locations of our synthetic apertures.  These apertures were overlaid
on the scaled \hi\ image, and the positions adjusted slightly when
required to center the apertures within the \hi\ holes.  This was
necessary since the reported accuracy of the P92 positions was $\pm$0.1\arcmin. 
Next, we divided the holes into four categories according to
size (again based on the data from P92), and assigned aperture sizes
accordingly.  The apertures adopted for the four size groups are
listed in Table \ref{table:apsizes}.  The decision to use variable
aperture sizes for the photometry was motivated by several factors:
(1) the central positions of the holes become increasingly uncertain
for larger holes; (2) there is no {\it a priori} reason to expect the
putative star clusters to be precisely in the hole centers; (3) the 
larger holes tend to be older so that diffusion would tend to increase 
the size of the clusters in these holes.  The apertures used ensure that 
we did not miss the clusters.

Because substantial galaxian background is present in many hole
locations, we measured the flux in both a circular aperture
corresponding to the hole location and in a concentric annular ring
surrounding the aperture.  This allows us to compare the fluxes and
colors in the two regions.  The area of each annular region was chosen
to be approximately equal to that of the corresponding circular
aperture to ensure comparable signal-to-noise for the two
measurements.  The test for the presence of a star cluster within the
\hi\ holes then involves comparing the fluxes within each
aperture/annulus pair to look for an excess of flux (expected to be
blue in color) coming from the central circular aperture.

In several instances, bright foreground stars were located inside the
aperture or annular regions we had marked for photometry.  In such
cases we either adjusted the positions of the hole centers slightly,
if possible, to avoid contamination from these objects, or excluded
those holes from the sample.  Photometry was successfully executed on
44 of the 51 hole regions identified in P92.  The locations of the
apertures and annuli are shown on the radio map (Figure~\ref{fig:hi
map}) and on the composite \bvr\ optical image
(Figures~\ref{fig:combined BVR image}).  Results from photometry of
the aperture/annular regions are given in Table~\ref{table:photholes}.
The equatorial coordinates of our aperture centers are given for each
hole, along with the aperture (inner diameter) and annular (outer
diameter) sizes in arc seconds, plus the $B$ band magnitude, $B$$-$$V$
and $B$$-$$R$ colors, and average $B$-band surface brightness of the
integrated light within both the aperture and annulus.  Throughout
this paper we use 1950 coordinates, for consistency with P92.  Note
that the background-subtracted flux measured in the annulus of hole
\#15, as well as for both the aperture and annulus of hole \#27, was
slightly negative, resulting in indeterminate magnitudes and colors.
These two holes contain no measurable light from either foreground
objects or from Ho$\thinspace$II itself.

\subsubsection{Point Sources in the HI Hole Regions}

A number of the holes in Ho$\thinspace$II had faint objects which appeared 
within the region delimited by our circular apertures or in the surrounding
annulus.  Since these objects could be the remnant star clusters which
produced the sequential SN explosions described in P92, we wished to
measure accurate magnitudes and colors for them.  Photometry was
performed on 29 such objects using a 5\arcsec\ diameter aperture and
a background annulus with an inner diameter of 8\arcsec\ and an outer
diameter of 20\arcsec.

Because we were measuring magnitudes and colors of individual sources
rather than an entire hole region, we decided to use the \bvr\ images
from the April 1995 observing run alone for this step, rather than
using the images merged from both observing runs.  The April 1995 data
were of better image quality compared to those from February 1994, and
this improvement in resolution almost completely compensated for the
added depth of the combined images with regard to point-source
detection.  The resolution (PSF FWHM) of the April 1995 broad-band
images is $1.7-1.8$\arcsec, and the 4$\sigma$ limit on the brightness
of a point source is $B = 23$.  Results from the point source
photometry are given in Table~\ref{table:photclusters}.  The first
column lists an object designation reflecting the hole number in which
the object is located, the next two columns give the position of the
object, and the $B$ magnitude and colors of the source are given in
columns 4 -- 6.  We note that the flux from {\it any} object within
the apertures or annuli was measured, regardless of the object's
brightness or location.  Therefore, this list should not be taken as a
list of possible candidate star clusters.  In many cases the objects
listed have magnitudes and colors consistent with their being faint
foreground stars.

\subsection{Results of Photometry}
\label{section:photometry results}

We summarize the results of our imaging and photometric analysis in
Table~\ref{table:photsummary}.  In this table, each of the 51 \hi\ holes
is characterized with regard to the following key question: is there
evidence for a star cluster at or near the center of the \hi\ hole?  The
categories into which each hole has been assigned are: (1) Empty Hole -- No
source within the central aperture exceeding a flux level of 3$\sigma$
above the annular flux level.  (2) Galaxian Background -- Source(s)
present in the hole, but with the characteristics/appearance of
general galaxian background light rather than the putative star
clusters.  (3) Possible Star Cluster -- Source(s) within the hole with
the correct characteristics (color, brightness, appearance) of a
genuine star cluster.  (4) Possible Photoionization Region -- The hole 
is coincident with an \hii\ region and is most likely a cavity of
photoionized gas rather than a wind/SNe-blown hole.  (5) Faint Foreground 
Star -- Photometry reveals that the object in the hole has the brightness 
and colors consistent with it being a foreground star not associated with 
Ho II.  (6) Contaminated/No Photometry -- A bright foreground star is 
present in the hole; no photometry was attempted.

It should be stressed that the assignment of a given hole into one of
the six categories is by no means unambiguous in all cases.  For
example, assigning objects to category 2 (galaxian background) as
opposed to category 5 (faint foreground star) was at times subjective.
Further, we emphasize that objects in category 3 (possible star
cluster) are assigned that designation even if the color and
brightness of the source is only {\it broadly} consistent with the
expected values.  An object classified as category 3 cannot be
interpreted as having been shown to be a young star cluster within
Ho$\thinspace$II.  Rather, it is {\it consistent} with that
hypothesis, but could just as likely be a foreground star (e.g., a
halo white dwarf).  Two holes (numbers 1 and 8) have point source
photometry listed in Table~\ref{table:photclusters}, but the sources
in question are located in the outer annuli.  Hence, these two holes
are classified as category 1.

The main results from Table~\ref{table:photsummary} that we want to
emphasize are: (1) nearly one third of the holes have no obvious
optical source within our photometric apertures, and (2) only a
minority of holes (6 of 44 for which photometry was obtained) have
optical sources with the colors and brightnesses consistent with the
expected star cluster.  One hole (\#43) is located in a complex of
bright \hii\ regions, and may represent a hole which is created by
photoionization of the \hi\ gas by the O and B stars.  The status of
the remaining holes (category 2 and 5) are uncertain, owing to the
foreground or galaxian light they contain.  However, in most cases
this light is inconsistent with the properties of the putative star
clusters, as quantified in the following section.

We note that the photometry listed in Tables~\ref{table:photholes} and
\ref{table:photclusters} has not been corrected for reddening, either
due to the Milky Way or internal to Ho$\thinspace$II.  Burstein \&
Heiles (1984) list a color excess of $E$($B$$-$$V$) $=$ 0.03 due to
foreground reddening in the direction of Ho$\thinspace$II.  With
regard to intrinsic absorption in Ho$\thinspace$II, spectra of four
\hii\ regions by Hunter \& Gallagher (1985) yield Balmer decrements
(H$\alpha$/H$\beta$) consistent with zero or modest reddening.  Hence,
we believe that dust is not significantly affecting our photometry.
This should be especially true in the outermost regions of
Ho$\thinspace$II, where our photometry provides the most sensitive
test of the SNe hypothesis.

We will return to the photometry results in more detail in 
section~\ref{section:discussion}.

\section{Modeling Analysis}
\label{section:analysis}

P92 catalogued the observed properties of the \hi\ holes in
Ho$\thinspace$II, and from these, derived quantities such as the ages
of the holes and the kinetic energies required to create them.  Some
of these observed and derived quantities (such as the radial expansion
of the holes, and their energy requirements) were interpreted as
evidence in favor of the stellar wind/multiple SNe scenario for the
origin of the holes.  In order to compare directly our observations
with predictions that arise from the SN scenario, we have used the
hole properties tabulated in P92 to derive the observable
characteristics of the clusters which should exist if the SN
hypothesis is correct.

A number of the quantities tabulated in P92 are distance-dependent;
for example, the kinetic energies they calculated are given by the
equation (\cite{chev74}):
\begin{equation}
\label{equation:energy}
E = 5.3 \times 10^{43}\ n_{HI}^{1.12}\ R^{3.12}\ V^{1.4}\ {\rm ergs}
\end{equation}

\noindent where $n_{HI}$ is the volume density of the surrounding
medium in particles per cubic centimeter, $R$ is the radius of the
hole in parsecs, and $V$ is the expansion velocity of the hole in
\kms.  P92 adopted a distance of 3.2 Mpc for Ho$\thinspace$II.  A more
recent distance determination has been done by Hoessel et al.~(1998)
using Cepheid variable stars in Ho$\thinspace$II.  We have adopted
their revised distance modulus for our calculations: $(m-M) = 27.42$,
which yields a distance of 3.05 Mpc.  Therefore we have used the data
tabulated in P92 but have re-scaled it in accordance with this revised
distance.  Some of the relevant quantities that changed with the new
distance have been included in Table~\ref{table:energy.out}.  For most
holes, the revised distance resulted in computed hole energies that
are $\sim$14\%\ less than those tabulated by P92.  Note, however, that
a typographical error in P92 caused their published energy value for
hole \#43 to be underestimated by more than an order of magnitude.
Our table reflects the corrected energy for this hole.

To derive the characteristics that should be observable if the remnant
clusters are located inside the \hi\ holes, we proceeded as follows:
using the kinetic energies of the holes, the number of supernova
explosions required to create each hole was calculated.  The energy
imparted to the ISM by one supernova explosion was taken to be
10$^{51}$ ergs (cf.~\cite{mccray87}).  If the number of supernova
explosions required to create a hole was $<$1, that hole was omitted
from the rest of the calculations.  For each remaining hole, a
Salpeter IMF (\cite{salpeter55}) was used to calculate the mass
distribution for a model star cluster.  This distribution was then
scaled so that the number of stars of mass $\geq$ 7 $M_{\sun}$ was
equal to the number of supernovae required to create that particular
hole.  (Stars with masses $\gtrsim$ 7 $M_{\sun}$ were assumed to be of
sufficient mass to end their lives as Type II supernovae.)  Next,
stars which would have evolved off the main sequence over a time scale
equal to the age of the hole were removed from the distribution.
Finally, composite magnitudes and colors were calculated for each of
the synthetic clusters, for comparison with our observations.

The same calculations were repeated for a Miller-Scalo IMF
(\cite{miller79}) and using a limit of $\geq$ 8 $M_{\sun}$ for the
lowest mass stars which end their lives as Type II supernovae.
Performing the calculation with these two sets of parameters provides
us with a reasonable range of predicted brightnesses for the putative
clusters, and reflects the current levels of uncertainty in both the
slope of the IMF and lower mass limit for Type II supernova
precursors.

We note that in this simple calculation we make no effort to take into
account the light that the post-main-sequence stars would contribute
to the cluster total.  For clusters in the age range considered here,
red supergiants might make a significant contribution to the total
light.  By ignoring the evolved stars, our models are providing only
{\it lower limits} to the total cluster brightness.  This must be kept
in mind when comparisons are made with the observations.

The final results of the modeling analysis are given in
Table~\ref{table:energy.out}.  The hole energies and ages listed in
columns 2 and 3 differ slightly from those tabulated by P92, for the
reasons mentioned above.  The predicted number of SNe for each hole is
one-tenth of the number given in column 2 (i.e., the hole energy
divided by 10$^{51}$).  Only those holes with energies in excess of
10$^{51}$ ergs (i.e., those which require at least one SN) are listed
in the table.  The $B$ magnitude and $B$$-$$V$ color listed under the
heading ``Model I'' are the composite values for the model clusters
computed using the Salpeter IMF and a Type II SN mass limit of 7
$M_{\sun}$.  The corresponding values listed under ``Model II'' are
for models using a Miller-Scalo IMF and a mass limit of 8 $M_{\sun}$.
In general, Model II predicts brighter clusters.

For many of the holes, the predicted
cluster brightnesses are well below our observational limits, and
hence we are unable to say anything definitive regarding the presence
or absence of a young stellar population.  However, for holes requiring
energies in excess of 10$^{52}$ ergs (or $>$ 10 SNe) the expected
brightness of the putative star cluster is at or above the limits set
by our data.  These holes will receive careful scrutiny in the next
section. 

\section{Discussion}
\label{section:discussion}

\subsection{Results}
\label{section:results}

The main results of our study are summarized in
Table~\ref{table:comparison}, which combines data from
Tables~\ref{table:photholes}, \ref{table:photclusters},
\ref{table:photsummary}, and \ref{table:energy.out} for specific
holes.  In Section~\ref{section:observations} we established that a
point source with $B$ $=$ 23.0 would be detected at the 4$\sigma$
level in our images.  We list in Table~\ref{table:comparison} all \hi\
holes that are predicted to contain star clusters as bright or
brighter than $B$ $=$ 23.0 in one or both of the models presented in
Section~\ref{section:analysis}.  The entries are sorted by hole
category, as taken from Table~\ref{table:photsummary}.

A number of holes characterized in Table~\ref{table:photsummary} as
category 1 (no objects visible in the hole) are seen to have limiting
magnitudes significantly fainter than the expected brightness of the
putative clusters.  In other words, if the SNe hypothesis for the
origin of the holes is correct, one should definitely be able to see
the clusters in a number of cases.  Specific holes for which this test
fails include 10, 13, 47, 49, and 50.  In all these cases, the
observational limits are more than one magnitude fainter than the
predicted values for the case of Model II (8 M$_\odot$ mass limit for
SN).  If the light from the evolved cluster stars is taken into
account, the differences between the predicted brightnesses and the
observed limits becomes even greater.  It is interesting to note that
these five holes are among the most energetic in Ho$\thinspace$II,
each requiring between 32 and 65 supernovae to create them in the P92
scenario.

For the specific \hi\ holes mentioned above, we can clearly rule out
the presence of the predicted star clusters at the expected levels.
Therefore, it seems extremely unlikely that these holes were created
by multiple SNe as hypothesized by P92.  This is perhaps no great
surprise, since the holes in question are located in regions of
extremely low surface brightness.  In fact, most lie outside of the
Holmberg radius of Ho$\thinspace$II.  At these large distances from
the galaxian center, the \hi\ gas column densities are quite low, well
below the empirical limits suggested by Kennicutt (1989) necessary for
star formation to occur.  In general, little star formation should
occur beyond the Holmberg radius, so that the existence of multiple
massive star clusters in the outer regions of Ho II, as predicted by
P92, would be at odds with what we know about star formation in other
galaxies.

The limits we can place on the presence of star clusters in the inner
\hi\ holes are less interesting for two reasons.  First, the galaxian
background is much higher there, making it easier to hide the presence
of any cluster light.  Second, the inner holes tend to be smaller in
size and hence require fewer SNe to create them.  Consequently the
brightnesses predicted for these holes tend to be fainter than our
observational limits.  Therefore, we cannot rule out the possibility
that some of these inner \hi\ holes are produced by SNe.  There is of
course no compelling reason for believing that all the holes have the
same origin; we can only adequately test the SNe hypothesis for the
outer \hi\ holes.  A few holes for which the observations may actually
support the SNe hypothesis of P92 are numbers 21, 36, 44, and 48.
These are listed in Table~\ref{table:comparison}.  Holes 21 and 48 are
the most likely candidates, although hole 21 is in a very crowded
region with many point sources in the optical images.  The sources in 
holes 36 and 44 are both significantly brighter than the model 
predictions.

Further evidence against the SNe hypothesis comes from the H$\alpha$
image, which is shown in Figure 3 with the \hi\ hole apertures
superposed.  Given the predicted ages and energetics of the holes, one
might expect that at least some of the holes would exhibit diffuse
H$\alpha$ emission.  No such emission is seen in any of the holes,
with the possible exception of the few holes that are coincident with
\hii\ regions (e.g., holes 16, 20, and 43).  The locations of
H$\alpha$ emission trace out the regions of high \hi\ column density
seen in Figure 1, indicating that the current star formation is
occurring at local density maxima in the neutral gas distributions.
P92 interpreted this as SNe-induced star formation caused by the
compression of swept-up gas.  We note, however, that this
interpretation does not appear to be consistent with most of the
holes, since only the minority have H$\alpha$ emission nearby.

ROSAT observations of Ho$\thinspace$II have failed to detect the
presence of any diffuse X-ray gas inside the HI cavities (F.~Walter \&
J.~Kerp, private communication).  Such X-ray emission would be
expected if the holes were filled with hot coronal gas from the SNe
explosions.  Furthermore, Stewart et al.~(1997) analyzed
far-ultraviolet (FUV) images of Ho$\thinspace$II taken with the
Ultraviolet Imaging Telescope and found no bright FUV knots located
within the \hi\ hole locations identified by P92.  They found instead
that bright FUV emission, if present, was likely to occur outside the
hole boundaries.

\subsection{Alternative Explanations}
\label{section:alternatives}

The observational evidence presented here strongly suggests that at
least some of the \hi\ holes in Ho$\thinspace$II are not caused by
multiple SN explosions in the manner envisioned by P92.  Here we
discuss possible alternative explanations for the presence of these
features.

\subsubsection{Modified Hole Energetics}
\label{section:modified energetics}

One possibility is that the SNe hypothesis is correct, but that the
numbers published by P92 for the energetics of the \hi\ holes are
systematically overestimated.  The energies and ages derived by P92
depend critically on the observed expansion velocities of the holes.
These are difficult to measure precisely, and in some cases the
evidence for the expansion is weak at best.  If the measured expansion
velocities given in P92 are systematically too high, then the holes
might actually be substantially older, and the true required energies
would be significantly reduced.  This being the case, it might be
possible to reduce the expected number of SNe, and hence the
brightness of the remnant star clusters, below the observational
limits found from our data.  However, for the largest holes, a
reduction in the total number of SNe of more than a factor of $\sim$5
is required.  This would then imply an overestimation of the expansion
velocities by a factor of more than three.  Such a large error in the
expansion velocities seems unlikely.  Furthermore, the additional
factors mentioned above regarding star formation beyond the Holmberg
radius and the lack of H$\alpha$ and X-ray gas could still pose
problems for this interpretation.

Another possibility is that the energetics of the holes have been
overestimated not because of errors in the the expansion velocities,
but because of uncertainties in the energy calculation itself.
Estimates for the amount of energy imparted to the ISM by stellar
winds and supernovae are based on our knowledge of the efficiencies of
these processes, which are not well-constrained.  P92 calculated the
energy associated with each of the \hi\ holes in Ho$\thinspace$II
using an expression derived from a hydrodynamical model by Chevalier
(1974), which describes the evolution of a single
spherically-symmetric supernova remnant in a uniform medium.  It seems
possible that using the results from such a model to calculate the
total energies associated with \hi\ holes in the non-uniform ISM of a
galaxy like Ho$\thinspace$II could introduce uncertainties of at least
a factor of a few.  As explained in the previous paragraph, if the
energies of the holes have been overestimated, then fewer SNe may be
needed to create the \hi\ holes, which could push the putative cluster
brightnesses below our observational limits.  As before, however, this
explanation still requires that some of the \hi\ holes be created by
supernovae occurring well beyond the Holmberg radius, in regions of
the galaxy where star formation does not appear to have occurred in
the past.

\subsubsection{A Non-Standard IMF}
\label{section:non-standard IMF}

Another way to retain the SNe hypothesis would be to invoke an unusual
IMF for the putative clusters.  This possibility was actually
suggested by P92.  A top-heavy IMF, rich in massive stars but poor in
low-mass ones, could explain the observations.  However, there is as
yet no real evidence for significant variations in the IMF as measured
in different environments in the Milky Way and other nearby galaxies
(\cite{leitherer98}).  Furthermore, since the stars which would be
providing the expected cluster signature are themselves fairly
massive, the IMF slope required to produce the requisite number of SNe
would have to be quite severe, perhaps even inverted, in order to not
produce a detectable population of B and A stars.  Invoking such an
unusual IMF is not a very compelling explanation.

Since the SNe hypothesis has significant problems, even when
allowances for the above variants are made, we are forced to consider
the alternative that the outer \hi\ holes are not produced by the
action of stellar winds and SNe.  We again stress that our analysis
does not exclude the possibility that the inner holes {\it are}
produced by SNe/winds.

\subsubsection{Gamma-Ray Bursts}
\label{section:GRBs}

One possible explanation that has gained attention recently is the
suggestion that the holes are remnants of Gamma-Ray Burst events
(GRBs).  Recent work by Loeb \& Perna (1998) and Efremov et al.~(1998)
proposes that some of the HI supershells and hole features seen in
nearby galaxies (such as the dwarfs in our sample) are remnants of
GRBs.  These authors suggest that GRBs might be associated with the
release of gravitational binding energy during, for example, the
collapse of a single, massive star to a black hole.  They argue that
such an event could produce a blast wave with energy comparable to the
multiple-SN events thought to be necessary to produce the \hi\ hole
features.  Since this proposal does not necessarily require that a
large star cluster be left behind after the explosive event, the
observational test applied here does not rule it out.  Because only a
single star can account for the GRB, one could hypothesize that the
explosions which create the \hi\ holes do not occur in massive star
clusters, but rather in smaller associations which we would have no
hope of detecting in our data.  However, like the SN scenario, the GRB
scenario requires that massive stars be present at the centers of the
\hi\ holes, in order to produce the expanding blast wave that creates
the hole.  As we have noted, many of the holes in Ho$\thinspace$II
occur in extremely LSB regions which show no indication of recent
massive star formation.  In addition, the GRB hypothesis would still
predict the presence of hot X-ray emitting gas within the \hi\ cavity.
Hence, although the GRB hypothesis is an attractive alternative to the
multiple-SNe scenario, and may well explain some of the \hi\ holes
seen in Ho$\thinspace$II, we consider it unlikely to be correct for
the outer holes, i.e., the same ones for which our current study has
ruled out the SNe hypothesis.

\subsubsection{Impacts from High-Velocity Clouds}
\label{section:HVCs}

If stellar energy sources are ruled out, the next two most likely
explanations for the \hi\ holes are large-scale dynamical effects and
ionization.  The dynamical effects of a collision between an infalling
neutral gas cloud and a galactic disk, resulting in an \hi\ hole, were
modeled by Tenorio-Tagle (1980, 1981).  Clear associations between
\hi\ holes and high-velocity clouds are known in the Milky Way (Heiles
1985), M101 (van der Hulst \& Sancisi 1988; Kamphuis, Sancisi, \& van
der Hulst 1991), NGC 628 (Kamphuis \& Briggs 1992), NGC 6946 (Kamphuis
\& Sancisi 1993), and NGC 5668 (Schulman et al.~ 1996).  The
high-velocity clouds could be primordial material or the result of a
galactic fountain.

This mechanism is attractive for explaining the holes at large
galactocentric radii in Ho$\thinspace$II, where there is very little
starlight and very little likelihood for star formation.  One of us
(DJW) has re-examined the data cubes from P92 to search for candidate
high-velocity clouds.  The data cube with the greatest sensitivity,
with a synthesized beam of 28 $\times$ 27 arc seconds and pixels of 10
arc seconds, was examined in the velocity range 64 to 244 km s$^{-1}$.
The root-mean-square (RMS) signal in an apparently line-free channel
was 1.88 mJy beam$^{-1}$.  Candidate clouds were found by first making
a statistical search for bright pixels, then searching for extended
bright areas around those pixels.  A candidate was required to have
extended, superimposed signal in at least three adjacent channels.

Only one candidate cloud met our requirements.  It is located at
08$^h$ 16$^m$ 37$^s$ 70\arcdeg\ 42\arcmin\ 36\arcsec, with a central
velocity of 223.5 km s$^{-1}$ and a total velocity width of 7.7 km
s$^{-1}$. It is 17\arcmin\ from the center of Ho$\thinspace$II, giving
a projected separation of about 15 kpc at the assumed distance of 3.05
Mpc.  Pixels brighter than the background RMS are found in three
channels. The peak intensity is 9.81 mJy beam$^{-1}$ (5.2 times the
RMS) in the raw maps, and 19.63 mJy beam$^{-1}$ after correction for
the primary beam.  The maximum brightness temperature is 2.4 K.  The
cloud's integrated \hi\ line signal is 4.9 Jansky km s$^{-1}$, giving
a total \hi\ mass of 1.2 $\times$ 10$^7 M_{\sun}$.  Its total angular
extent is 92\arcsec\ or 1.4 kpc.  The properties of this candidate
cloud are similar to those of the Milky Way high-velocity clouds
described by Wakker \& van Woerden (1997).

This cloud is not easily visible in the data cube, so no mention was
made of it in P92.  We wish to emphasize that this is a marginal
detection at best, and must be confirmed by an independent observation
before it can be considered anything but a candidate cloud.  More than
two million apparently empty pixels were searched for local peaks
which might be candidate clouds. Only two pixels were found to be as
bright as 5.2 times the RMS, and of these, only one had extended
signal in three channels.  In a sample of this size, two pixels as
bright as 5.2 times the RMS are expected if the noise is Gaussian.
There is a good possibility that we have simply found a noise peak
rather than a real cloud.  Even so, the candidate cloud can be used to
set limits on the presence of high-velocity clouds within the observed
velocity range.

The cloud has a velocity of 65 km s$^{-1}$ relative to
Ho$\thinspace$II, which has a systemic velocity of 158 km s$^{-1}$.
Its kinetic energy in the frame of Ho$\thinspace$II is 5 $\times$
10$^{46}$ Joules or 5 $\times$ 10$^{53}$ ergs, large enough to cause
the largest holes in Ho$\thinspace$II.  The possible detection of a
single candidate cloud with enough energy to cause a hole does not
prove or disprove the hypothesis that infall caused the holes.  Infall
could be episodic, or there could be clouds outside the observed
velocity range.  It does show the observational difficulty in
identifying high-velocity clouds in galaxies at the distance of the
M81 group --- even in observations as deep as those of P92, candidate
clouds are nearly indistinguishable from 5-$\sigma$ noise peaks.

\subsubsection{Large-Scale Turbulence}
\label{section:turbulence}

Holes might be unavoidable due to the nature of the interstellar
medium.  It is well known that molecular clouds are fractal --- see
Beech (1987), Bazell \& D\'esert (1988), Falgarone (1989), Scalo
(1990), Falgarone, Phillips, \& Walker (1991), and Elmegreen \&
Falgarone (1996).  It is becoming clear that \hi\ is fractal as well
--- see Vogelaar \& Wakker (1994) and Westpfahl et al. (1999).  The
fractal dimension of the ISM is similar to that of structures seen in
laboratory turbulence, which has led Elmegreen \& Efremov (1999) and
others to conclude that interstellar clouds form by processes related
to turbulence.  The processes which cause fractal structure, including
turbulence, usually produce an internal distribution of holes,
characterized as lacunarity by Mandelbrot (1983).  If \hi\ clouds are
produced by processes related to turbulence, the holes may be a
manifestation of the formation process.

We note that the fractal nature of \hi\ distributions may change the
energy required to form a hole via supernovae and stellar winds.  A
fractal structure with significant lacunarity may provide natural
chimneys through which supernova ejecta can flow, thus significantly
increasing the amount of energy injection needed to form an expanding
shell.

\subsubsection{Ionization}
\label{section:ionization}

Holes might also be formed by ionization.  A source of ionizing
photons in the outer regions of Ho$\thinspace$II might be the
intergalactic UV field.  The observed column densities in the HI gas
at the locations of these outer holes is of order a few times
10$^{20}$ cm$^{-2}$, significantly above the densities at which the UV
radiation field can keep a large fraction of the HI ionized for an HI
disk of normal thickness.  However, P92 argue convincingly that the
scale height of the gas in Ho$\thinspace$II is significantly larger
than in a typical spiral disk.  If this is correct, the actual volume
density of the gas in the outer regions would be significantly less
than that for gas in a spiral disk with the same measured column
density.  In this case, it might be possible for lower density pockets
of gas to approach the threshold volume density below which the
ionization fraction of the gas remains high in steady state due to the
ionizing UV photons.  In other words, once a pocket of low density gas
is created in the outer parts of a puffed-up disk as is envisioned for
Ho$\thinspace$II, it could become ionized by the intergalactic UV
radiation field and remain highly ionized for a long time.  Eventually
the holes would be destroyed by dynamical processes (e.g., rotational
shear), but in the outer parts of the galaxy they might well maintain
their integrity for a substantial length of time due to the slow
rotation speeds in low-mass galaxies like Ho$\thinspace$II.

\subsubsection{An Unresolved Question}
\label{section:unresolved question}

The actual origin of the \hi\ holes in the outer parts of
Ho$\thinspace$II remains an open question.  In the present study, we
suggest strongly that at least some of the holes are not produced by
the combined action of stellar winds and SNe explosions.  In many ways
the SNe hypothesis is a natural and sensible explanation for the
holes, which was perhaps why it was almost universally accepted on
face value.  However, in several instances it clearly fails the direct
observational test which we have applied.  Additional work needs to be
done on this problem in order for a clearer picture of the nature of
these large-scale features to be developed.  Our current lack of
understanding leaves open a number of questions regarding the
evolution of the ISM in irregular galaxies, and in particular the
actual role of feedback from massive stars in shaping the ISM.

\section{Summary \& Conclusions}
\label{section:summary}

We have carried out a deep, multi-color imaging study of
Ho$\thinspace$II, a dwarf galaxy in the M81 group which has been shown
to contain a large number of expanding holes in its neutral hydrogen
distribution.  The formation of the \hi\ holes in Ho$\thinspace$II and
other galaxies like it has been attributed to multiple SNe occurring
within wind-blown shells around young, massive star clusters.  To
search for evidence of the clusters, we have compared our optical
images with the published \hi\ maps, and have measured accurate
magnitudes and colors of all objects in and around the \hi\ holes.
Photometry of 44 hole regions in Ho$\thinspace$II reveals that at
least 16 holes contain no detectable point sources brighter than $B$
$=$ 23.0.  Ten of these holes are located beyond the Holmberg
radius. An additional 21 holes contain only red ($B$$-$$V$ $>$ 1.0)
sources, which are most likely either faint foreground stars or
diffuse galaxian background emission from Ho$\thinspace$II.  Only 6
holes contain sources which could be interpreted as being young
clusters of stars with the requisite brightness and color.

Comparison of models which predict the brightness of the putative star
clusters with the observational limits obtained from our imaging data
appear to rule out the SN scenario as being the cause of at least
several of the most substantial \hi\ holes.  While convincing
arguments cannot be made against the SNe scenario for the majority of
the holes due to the lack of depth of our images coupled with severe
crowding in the central portions of the galaxy, the fact that at least
several holes appear to require an alternative explanation for their
origin raises doubts about the SNe scenario in general.  The lack of
diffuse H$\alpha$ and X-ray emission from any of the holes further
supports the possibility that the SNe scenario may be incorrect.
Recent suggestions that the \hi\ holes in galaxies like
Ho$\thinspace$II are caused by the events which create Gamma-ray
bursts are also not favored by the current findings, although such
scenarios are more difficult to rule out with the optical data since
they require only a single massive star.  A number of other
alternative explanations for the existence of the \hi\ holes are
explored, including errors in the hole energetics, non-standard IMFs,
dynamical processes such as large-scale turbulence or impacts from
high-velocity clouds, and ionization.  None of these alternatives is
clearly favored at this time, and the origin of the \hi\ holes remains
an open question.

There is no doubt that energy input from massive stars plays a major
role in shaping the ISM in galaxies.  The current study, however,
suggests that one must interpret the observational evidence for such
influence carefully.  Although the scenario proposed by P92 appears
sensible, it makes a direct observational prediction which is not
verified by the current study.  The precise role that winds from
massive stars and SNe shocks play in sculpting the gaseous
distribution in galaxies remains an open question, calling for
continued careful work on both the observational and theoretical
fronts.

\acknowledgments

We are grateful for the professional support of the staff of Kitt Peak
National Observatory during our two observing trips.  This research
has made use of the NASA/IPAC Extragalactic Database (NED) which is
operated by the Jet Propulsion Laboratory, California Institute of
Technology, under contract with the National Aeronautics and Space
Administration.  KLR, JJS and LAR acknowledge with gratitude financial
support from Wesleyan University, the Keck Northeast Astronomy
Consortium, the Research Corporation, and the National Science
Foundation, all of whom provided partial support for this project.
DJW gratefully acknowledges financial support from the New Mexico
Space Grant Consortium, the Kahlmeyer Foundation, and the Research and
Economic Development division of New Mexico Tech.  We have benefited
from discussions with many colleagues, including D. Puche, R. Larson,
S. Van Dyk, F. Walter, and E. Brinks.  JJS would like to recognize
L. van Zee and NRAO-Socorro for their hospitality during the
preparation of portions of this paper.  We would also like to thank
the anonymous referee for useful comments.

\clearpage

%
%

\clearpage
\begin{deluxetable}{ccc}
\tablecaption{Aperture Sizes Used for Photometry of HI Hole Regions
\label{table:apsizes}}
\tablewidth{300pt}
\tablenum{1}
\tablehead{
\colhead{Aperture Size} & \colhead{Inner Diameter (\arcsec)} & \colhead{Outer Diameter (\arcsec)}\\
\colhead{(1)} & \colhead{(2)} & \colhead{(3)}
}
\startdata
Tiny & 11.4 & 16.0\nl
Small & 14.2 & 20.0\nl
Medium & 17.0 & 24.0\nl
Large & 21.2 & 30.0\nl
\enddata
\tablecomments{Column 2 lists
diameters of the circular apertures, which also serve as the inner
diameters of the associated annuli.  Column 3 lists outer diameters
of the annuli.}
\end{deluxetable}

\clearpage
\begin{deluxetable}{crrccccr}
\tablewidth{520pt}
\tablenum{2}
\tablecaption{Photometry of Hole/Inter-Hole Regions \label{table:photholes}}
\tablehead{
\colhead{Region} & \colhead{$\alpha$ (1950)} &
\colhead{$\delta$ (1950)} &
\colhead{Inner} & \colhead{B} & 
\colhead{B$-$V} & \colhead{{B$-$R}} & 
\colhead{$\mu_{\rm{B}}$} \\
\colhead{} & \colhead{} & \colhead{} &
\colhead{Outer} & \colhead{} &
\colhead{} & \colhead{} & 
\colhead{} 
}
\startdata
 1 & 8:13:05.7 & +  70:52:19 & 21.20\arcsec & 22.376 $\pm$ 0.114 &  0.546 $\pm$ 0.198 &  0.460 $\pm$ 0.147 & 28.746 \nl 
 & & & 30.00\arcsec & 21.866 $\pm$ 0.129 &  0.883 $\pm$ 0.219 &  1.565 $\pm$ 0.162 & 28.238 \nl 
 & & & & & & & \nl 
 2 & 8:13:11.4 & +  70:53:47 & 21.20\arcsec & 20.566 $\pm$ 0.033 &  0.257 $\pm$ 0.055 &  0.511 $\pm$ 0.040 & 26.935 \nl 
 & & & 30.00\arcsec & 20.557 $\pm$ 0.037 &  0.176 $\pm$ 0.065 &  0.393 $\pm$ 0.045 & 26.929 \nl 
 & & & & & & & \nl 
 3 & 8:13:09.8 & +  70:53:15 & 21.20\arcsec & 20.954 $\pm$ 0.040 &  0.347 $\pm$ 0.068 &  0.715 $\pm$ 0.047 & 27.323 \nl 
 & & & 30.00\arcsec & 20.647 $\pm$ 0.045 &  0.341 $\pm$ 0.078 &  0.527 $\pm$ 0.053 & 27.019 \nl 
 & & & & & & & \nl 
 4 & 8:13:12.8 & +  70:56:39 & 14.20\arcsec & 22.536 $\pm$ 0.092 &  0.565 $\pm$ 0.155 &  1.354 $\pm$ 0.099 & 28.036 \nl 
 & & & 20.00\arcsec & 22.294 $\pm$ 0.108 &  0.526 $\pm$ 0.182 &  1.375 $\pm$ 0.116 & 27.775 \nl 
 & & & & & & & \nl 
 7 & 8:13:23.5 & +  70:54:14 & 11.40\arcsec & 21.931 $\pm$ 0.051 &  0.395 $\pm$ 0.086 &  1.088 $\pm$ 0.057 & 26.953 \nl 
 & & & 16.00\arcsec & 21.768 $\pm$ 0.060 &  0.500 $\pm$ 0.101 &  1.030 $\pm$ 0.067 & 26.757 \nl 
 & & & & & & & \nl 
 8 & 8:13:24.9 & +  70:56:08 & 17.00\arcsec & 21.099 $\pm$ 0.038 &  0.663 $\pm$ 0.055 &  1.348 $\pm$ 0.042 & 26.989 \nl 
 & & & 24.00\arcsec & 20.936 $\pm$ 0.043 &  0.222 $\pm$ 0.066 &  0.856 $\pm$ 0.048 & 26.818 \nl 
 & & & & & & & \nl 
 9 & 8:13:26.1 & +  70:55:07 & 21.20\arcsec & 20.982 $\pm$ 0.040 &  0.658 $\pm$ 0.059 &  1.452 $\pm$ 0.044 & 27.351 \nl 
 & & & 30.00\arcsec & 21.164 $\pm$ 0.047 &  0.254 $\pm$ 0.074 &  1.154 $\pm$ 0.052 & 27.536 \nl 
 & & & & & & & \nl 
 10 & 8:13:27.7 & +  70:50:59 & 21.20\arcsec & 19.928 $\pm$ 0.026 &  0.268 $\pm$ 0.038 &  0.779 $\pm$ 0.030 & 26.297 \nl 
 & & & 30.00\arcsec & 19.933 $\pm$ 0.029 &  0.374 $\pm$ 0.043 &  0.689 $\pm$ 0.033 & 26.305 \nl 
 & & & & & & & \nl 
 11 & 8:13:29.6 & +  70:57:07 & 14.20\arcsec & 22.100 $\pm$ 0.065 &  0.723 $\pm$ 0.098 &  1.643 $\pm$ 0.069 & 27.599 \nl 
 & & & 20.00\arcsec & 21.721 $\pm$ 0.075 &  0.753 $\pm$ 0.113 &  1.411 $\pm$ 0.079 & 27.202 \nl 
 & & & & & & & \nl 
 12 & 8:13:37.3 & +  70:53:02 & 14.20\arcsec & 19.009 $\pm$ 0.023 &  0.493 $\pm$ 0.028 &  0.853 $\pm$ 0.026 & 24.508 \nl 
 & & & 20.00\arcsec & 18.947 $\pm$ 0.024 &  0.438 $\pm$ 0.029 &  0.791 $\pm$ 0.027 & 24.428 \nl 
 & & & & & & & \nl 
 13 & 8:13:43.8 & +  70:48:13 & 21.20\arcsec & 21.096 $\pm$ 0.044 &  0.122 $\pm$ 0.087 &  0.557 $\pm$ 0.053 & 27.465 \nl 
 & & & 30.00\arcsec & 20.955 $\pm$ 0.051 &  0.557 $\pm$ 0.099 &  0.869 $\pm$ 0.061 & 27.327 \nl 
 & & & & & & & \nl 
 14 & 8:13:42.3 & +  70:52:55 & 14.20\arcsec & 18.339 $\pm$ 0.022 &  0.468 $\pm$ 0.027 &  0.833 $\pm$ 0.025 & 23.838 \nl 
 & & & 20.00\arcsec & 18.339 $\pm$ 0.023 &  0.500 $\pm$ 0.028 &  0.872 $\pm$ 0.025 & 23.820 \nl 
 & & & & & & & \nl 
 15 & 8:13:41.8 & +  70:57:30 & 14.20\arcsec & 24.737 $\pm$ 0.591 &  1.730 $\pm$ 0.668 &  3.387 $\pm$ 0.593 & 30.236 \nl 
 & & &  20.00\arcsec &  .......... &  ..........  &  .......... &  ..........\nl 
 & & & & & & & \nl 
 16 & 8:13:45.4 & +  70:52:22 & 14.20\arcsec & 17.212 $\pm$ 0.022 &  0.332 $\pm$ 0.026 &  0.649 $\pm$ 0.024 & 22.711 \nl 
 & & & 20.00\arcsec & 17.324 $\pm$ 0.022 &  0.311 $\pm$ 0.026 &  0.589 $\pm$ 0.025 & 22.805 \nl 
 & & & & & & & \nl 
 18 & 8:13:45.2 & +  70:55:22 & 11.40\arcsec & 20.643 $\pm$ 0.028 &  0.369 $\pm$ 0.040 &  0.732 $\pm$ 0.032 & 25.666 \nl 
 & & & 16.00\arcsec & 20.591 $\pm$ 0.031 &  0.377 $\pm$ 0.045 &  0.799 $\pm$ 0.036 & 25.580 \nl 
 & & & & & & & \nl 
 19 & 8:13:46.0 & +  70:54:28 & 21.20\arcsec & 18.330 $\pm$ 0.023 &  0.389 $\pm$ 0.027 &  0.849 $\pm$ 0.025 & 24.699 \nl 
 & & & 30.00\arcsec & 18.378 $\pm$ 0.023 &  0.329 $\pm$ 0.028 &  0.664 $\pm$ 0.026 & 24.750 \nl 
 & & & & & & & \nl 
 20 & 8:13:47.3 & +  70:52:14 & 14.20\arcsec & 17.468 $\pm$ 0.022 &  0.377 $\pm$ 0.026 &  0.682 $\pm$ 0.024 & 22.967 \nl 
 & & & 20.00\arcsec & 17.538 $\pm$ 0.022 &  0.433 $\pm$ 0.027 &  0.793 $\pm$ 0.025 & 23.019 \nl 
 & & & & & & & \nl 
 21 & 8:13:49.4 & +  70:50:43 & 21.20\arcsec & 17.508 $\pm$ 0.022 &  0.162 $\pm$ 0.026 &  0.327 $\pm$ 0.024 & 23.877 \nl 
 & & & 30.00\arcsec & 17.670 $\pm$ 0.022 &  0.197 $\pm$ 0.027 &  0.408 $\pm$ 0.025 & 24.042 \nl 
 & & & & & & & \nl 
 22 & 8:13:51.5 & +  70:52:38 & 14.20\arcsec & 17.294 $\pm$ 0.022 &  0.410 $\pm$ 0.026 &  0.764 $\pm$ 0.024 & 22.793 \nl 
 & & & 20.00\arcsec & 17.318 $\pm$ 0.022 &  0.436 $\pm$ 0.027 &  0.801 $\pm$ 0.025 & 22.799 \nl 
 & & & & & & & \nl 
 23 & 8:13:51.9 & +  70:53:02 & 14.20\arcsec & 17.792 $\pm$ 0.022 &  0.464 $\pm$ 0.027 &  0.823 $\pm$ 0.025 & 23.291 \nl 
 & & & 20.00\arcsec & 17.777 $\pm$ 0.022 &  0.461 $\pm$ 0.027 &  0.813 $\pm$ 0.025 & 23.258 \nl 
 & & & & & & & \nl 
 24 & 8:13:54.6 & +  70:53:36 & 14.20\arcsec & 18.109 $\pm$ 0.022 &  0.470 $\pm$ 0.027 &  0.848 $\pm$ 0.025 & 23.608 \nl 
 & & & 20.00\arcsec & 18.092 $\pm$ 0.023 &  0.465 $\pm$ 0.027 &  0.831 $\pm$ 0.025 & 23.573 \nl 
 & & & & & & & \nl 
 27\tablenotemark{1} & 8:19:12.7 & +70:38:09 & 21.20\arcsec & .......... & .......... & .......... & ..........  \nl 
 & & &  30.00\arcsec &  .......... &  ..........  &  .......... &  ..........\nl 
 & & & & & & & \nl 
 28 & 8:13:59.1 & +  70:53:14 & 14.20\arcsec & 17.926 $\pm$ 0.022 &  0.444 $\pm$ 0.027 &  0.817 $\pm$ 0.025 & 23.426 \nl 
 & & & 20.00\arcsec & 17.950 $\pm$ 0.022 &  0.442 $\pm$ 0.027 &  0.815 $\pm$ 0.025 & 23.432 \nl 
 & & & & & & & \nl 
 30 & 8:14:00.9 & +  70:53:55 & 14.20\arcsec & 18.589 $\pm$ 0.023 &  0.498 $\pm$ 0.028 &  0.890 $\pm$ 0.025 & 24.088 \nl 
 & & & 20.00\arcsec & 18.557 $\pm$ 0.023 &  0.529 $\pm$ 0.028 &  0.949 $\pm$ 0.026 & 24.038 \nl 
 & & & & & & & \nl 
 31 & 8:14:01.3 & +  70:53:27 & 14.20\arcsec & 18.095 $\pm$ 0.022 &  0.409 $\pm$ 0.027 &  0.737 $\pm$ 0.025 & 23.594 \nl 
 & & & 20.00\arcsec & 18.138 $\pm$ 0.023 &  0.421 $\pm$ 0.027 &  0.763 $\pm$ 0.025 & 23.619 \nl 
 & & & & & & & \nl 
 32 & 8:13:59.4 & +  70:51:16 & 17.00\arcsec & 18.452 $\pm$ 0.022 &  0.361 $\pm$ 0.027 &  0.658 $\pm$ 0.025 & 24.342 \nl 
 & & & 24.00\arcsec & 18.483 $\pm$ 0.023 &  0.425 $\pm$ 0.028 &  0.773 $\pm$ 0.026 & 24.366 \nl 
 & & & & & & & \nl 
 33 & 8:14:03.8 & +  70:54:33 & 14.20\arcsec & 19.058 $\pm$ 0.023 &  0.345 $\pm$ 0.029 &  0.664 $\pm$ 0.026 & 24.557 \nl 
 & & & 20.00\arcsec & 19.053 $\pm$ 0.023 &  0.385 $\pm$ 0.030 &  0.734 $\pm$ 0.026 & 24.535 \nl 
 & & & & & & & \nl 
 34 & 8:14:03.8 & +  70:55:44 & 14.20\arcsec & 19.668 $\pm$ 0.024 &  0.437 $\pm$ 0.031 &  0.801 $\pm$ 0.028 & 25.167 \nl 
 & & & 20.00\arcsec & 19.622 $\pm$ 0.026 &  0.348 $\pm$ 0.033 &  0.693 $\pm$ 0.029 & 25.103 \nl 
 & & & & & & & \nl 
 35 & 8:14:04.2 & +  70:56:22 & 21.20\arcsec & 19.214 $\pm$ 0.024 &  0.621 $\pm$ 0.030 &  1.198 $\pm$ 0.027 & 25.584 \nl 
 & & & 30.00\arcsec & 18.851 $\pm$ 0.026 &  0.837 $\pm$ 0.031 &  1.414 $\pm$ 0.028 & 25.224 \nl 
 & & & & & & & \nl 
 36 & 8:14:06.6 & +  70:52:44 & 17.00\arcsec & 17.668 $\pm$ 0.022 &  0.385 $\pm$ 0.027 &  0.713 $\pm$ 0.025 & 23.557 \nl 
 & & & 24.00\arcsec & 17.547 $\pm$ 0.022 &  0.349 $\pm$ 0.027 &  0.659 $\pm$ 0.025 & 23.430 \nl 
 & & & & & & & \nl 
 37 & 8:14:06.3 & +  70:54:55 & 14.20\arcsec & 19.251 $\pm$ 0.023 &  0.365 $\pm$ 0.029 &  0.750 $\pm$ 0.026 & 24.750 \nl 
 & & & 20.00\arcsec & 19.093 $\pm$ 0.024 &  0.380 $\pm$ 0.030 &  0.718 $\pm$ 0.027 & 24.575 \nl 
 & & & & & & & \nl 
 39 & 8:14:09.9 & +  70:54:14 & 14.20\arcsec & 18.650 $\pm$ 0.022 &  0.281 $\pm$ 0.027 &  0.568 $\pm$ 0.025 & 24.149 \nl 
 & & & 20.00\arcsec & 18.679 $\pm$ 0.023 &  0.299 $\pm$ 0.028 &  0.607 $\pm$ 0.026 & 24.160 \nl 
 & & & & & & & \nl 
 40 & 8:14:11.0 & +  70:50:10 & 14.20\arcsec & 20.322 $\pm$ 0.027 &  0.116 $\pm$ 0.041 &  0.279 $\pm$ 0.033 & 25.821 \nl 
 & & & 20.00\arcsec & 20.747 $\pm$ 0.030 &  0.193 $\pm$ 0.048 &  0.391 $\pm$ 0.037 & 26.228 \nl 
 & & & & & & & \nl 
 41 & 8:14:12.1 & +  70:50:44 & 14.20\arcsec & 20.228 $\pm$ 0.026 &  0.294 $\pm$ 0.037 &  0.735 $\pm$ 0.030 & 25.728 \nl 
 & & & 20.00\arcsec & 20.297 $\pm$ 0.029 &  0.202 $\pm$ 0.041 &  0.494 $\pm$ 0.033 & 25.779 \nl 
 & & & & & & & \nl 
 42 & 8:14:09.3 & +  70:54:44 & 14.20\arcsec & 19.043 $\pm$ 0.023 &  0.322 $\pm$ 0.029 &  0.716 $\pm$ 0.026 & 24.542 \nl 
 & & & 20.00\arcsec & 19.119 $\pm$ 0.023 &  0.317 $\pm$ 0.030 &  0.664 $\pm$ 0.026 & 24.600 \nl 
 & & & & & & & \nl 
 43 & 8:14:13.5 & +  70:51:39 & 11.40\arcsec & 19.246 $\pm$ 0.023 &  0.309 $\pm$ 0.028 &  0.608 $\pm$ 0.026 & 24.269 \nl 
 & & & 16.00\arcsec & 18.874 $\pm$ 0.023 &  0.277 $\pm$ 0.029 &  0.500 $\pm$ 0.027 & 23.863 \nl 
 & & & & & & & \nl 
 44 & 8:14:12.1 & +  70:52:56 & 21.20\arcsec & 17.693 $\pm$ 0.022 &  0.382 $\pm$ 0.027 &  0.724 $\pm$ 0.025 & 24.063 \nl 
 & & & 30.00\arcsec & 17.780 $\pm$ 0.022 &  0.380 $\pm$ 0.027 &  0.703 $\pm$ 0.025 & 24.152 \nl 
 & & & & & & & \nl 
 45 & 8:14:12.7 & +  70:54:60 & 14.20\arcsec & 19.373 $\pm$ 0.023 &  0.247 $\pm$ 0.030 &  0.591 $\pm$ 0.027 & 24.873 \nl 
 & & & 20.00\arcsec & 19.383 $\pm$ 0.024 &  0.295 $\pm$ 0.031 &  0.673 $\pm$ 0.028 & 24.864 \nl 
 & & & & & & & \nl 
 46 & 8:14:14.7 & +  70:50:30 & 14.20\arcsec & 20.233 $\pm$ 0.026 &  0.355 $\pm$ 0.036 &  0.849 $\pm$ 0.030 & 25.732 \nl 
 & & & 20.00\arcsec & 20.464 $\pm$ 0.029 &  0.172 $\pm$ 0.041 &  0.644 $\pm$ 0.033 & 25.946 \nl 
 & & & & & & & \nl 
 47 & 8:14:16.8 & +  70:49:21 & 21.20\arcsec & 21.191 $\pm$ 0.047 &  0.041 $\pm$ 0.099 &  0.659 $\pm$ 0.056 & 27.560 \nl 
 & & & 30.00\arcsec & 20.954 $\pm$ 0.054 &  0.391 $\pm$ 0.112 &  0.976 $\pm$ 0.064 & 27.326 \nl 
 & & & & & & & \nl 
 48 & 8:14:22.9 & +  70:54:30 & 21.20\arcsec & 18.473 $\pm$ 0.022 &  0.189 $\pm$ 0.028 &  0.492 $\pm$ 0.025 & 24.843 \nl 
 & & & 30.00\arcsec & 18.365 $\pm$ 0.023 &  0.176 $\pm$ 0.028 &  0.441 $\pm$ 0.026 & 24.737 \nl 
 & & & & & & & \nl 
 49 & 8:14:24.2 & +  70:55:55 & 21.20\arcsec & 20.388 $\pm$ 0.030 &  0.233 $\pm$ 0.049 &  0.884 $\pm$ 0.035 & 26.758 \nl 
 & & & 30.00\arcsec & 20.406 $\pm$ 0.034 &  0.347 $\pm$ 0.057 &  0.947 $\pm$ 0.039 & 26.778 \nl 
 & & & & & & & \nl 
 50 & 8:14:33.1 & +  70:52:15 & 21.20\arcsec & 20.307 $\pm$ 0.029 &  0.170 $\pm$ 0.048 &  0.840 $\pm$ 0.034 & 26.677 \nl 
 & & & 30.00\arcsec & 20.301 $\pm$ 0.032 &  0.306 $\pm$ 0.055 &  0.844 $\pm$ 0.037 & 26.673 \nl 
 & & & & & & & \nl 
 51 & 8:14:33.2 & +  70:55:15 & 21.20\arcsec & 20.324 $\pm$ 0.031 &  0.824 $\pm$ 0.040 &  1.703 $\pm$ 0.033 & 26.693 \nl 
 & & & 30.00\arcsec & 20.459 $\pm$ 0.033 &  0.368 $\pm$ 0.045 &  0.911 $\pm$ 0.036 & 26.831 \nl 
\enddata
\tablenotetext{1}{Photometry of the annular region for Hole 15, and of
both the hole and annular region for Hole 27, formally yielded
negative counts after background subtraction.}
\end{deluxetable}

\clearpage
\begin{deluxetable}{crrccr}
\tablewidth{460pt}
\tablenum{3}
\tablecaption{Photometry of Point Sources in \hi\ Hole Regions  \label{table:photclusters}}
\tablehead{
\colhead{Object} & \colhead{$\alpha$ (1950)} &
\colhead{$\delta$ (1950)} & \colhead{B} & 
\colhead{B$-$V} & \colhead{{B$-$R}}
}
\startdata
 H1-1 & 8:13:05.5 & +  70:52:33 & 23.766 $\pm$ 0.214 &  1.662 $\pm$ 0.239 &  2.909 $\pm$ 0.220 \nl 
 & & & & & \nl 
 H8-1 & 8:13:27.0 & +  70:56:09 & 22.468 $\pm$ 0.087 & -0.101 $\pm$ 0.209 & -0.362 $\pm$ 0.296 \nl 
 & & & & & \nl 
 H9-1 & 8:13:27.2 & +  70:55:07 & 24.576 $\pm$ 0.450 &  2.491 $\pm$ 0.463 &  3.573 $\pm$ 0.454 \nl 
 & & & & & \nl 
 H12-1 & 8:13:37.6 & +  70:53:08 & 22.257 $\pm$ 0.086 &  0.578 $\pm$ 0.136 &  0.940 $\pm$ 0.133 \nl 
 & & & & & \nl 
 H16-1 & 8:13:46.2 & +  70:52:26 & 19.487 $\pm$ 0.033 & -0.013 $\pm$ 0.048 &  0.174 $\pm$ 0.046 \nl 
 & & & & & \nl 
 H19-1 & 8:13:45.8 & +  70:54:28 & 22.335 $\pm$ 0.101 &  1.305 $\pm$ 0.117 &  2.467 $\pm$ 0.106 \nl 
 H19-2 & 8:13:44.4 & +  70:54:25 & 21.472 $\pm$ 0.061 & -0.198 $\pm$ 0.120 & -0.136 $\pm$ 0.145 \nl 
 & & & & & \nl 
 H21-1 & 8:13:50.0 & +  70:50:44 & 21.227 $\pm$ 0.073 &  0.029 $\pm$ 0.116 & -0.068 $\pm$ 0.129 \nl 
 H21-2 & 8:13:49.7 & +  70:50:37 & 20.898 $\pm$ 0.050 & -0.089 $\pm$ 0.086 & -0.115 $\pm$ 0.096 \nl 
 H21-3 & 8:13:48.1 & +  70:50:55 & 21.342 $\pm$ 0.072 &  0.340 $\pm$ 0.101 &  0.933 $\pm$ 0.090 \nl 
 & & & & & \nl 
 H22-1 & 8:13:50.3 & +  70:52:44 & 21.530 $\pm$ 0.139 &  1.624 $\pm$ 0.146 &  2.516 $\pm$ 0.142 \nl 
 H22-2 & 8:13:52.3 & +  70:52:41 & 20.226 $\pm$ 0.044 & -0.181 $\pm$ 0.077 & -0.234 $\pm$ 0.097 \nl 
 H22-3 & 8:13:50.0 & +  70:52:33 & 20.786 $\pm$ 0.072 &  0.072 $\pm$ 0.113 &  0.374 $\pm$ 0.106 \nl 
 H22-4 & 8:13:48.7 & +  70:52:35 & 20.196 $\pm$ 0.048 & -0.322 $\pm$ 0.089 & -0.335 $\pm$ 0.102 \nl 
 & & & & & \nl 
 H23-1 & 8:13:49.3 & +  70:52:55 & 19.894 $\pm$ 0.033 &  1.051 $\pm$ 0.038 &  1.710 $\pm$ 0.036 \nl 
 & & & & & \nl 
 H30-1 & 8:13:59.3 & +  70:53:51 & 21.754 $\pm$ 0.064 &  1.171 $\pm$ 0.075 &  1.940 $\pm$ 0.070 \nl 
 & & & & & \nl 
 H32-1 & 8:13:58.9 & +  70:51:25 & 23.802 $\pm$ 0.396 &  1.901 $\pm$ 0.415 &  3.037 $\pm$ 0.401 \nl 
 & & & & & \nl 
 H35-1 & 8:14:02.9 & +  70:56:12 & 20.387 $\pm$ 0.034 &  1.594 $\pm$ 0.038 &  2.379 $\pm$ 0.036 \nl 
 & & & & & \nl 
 H36-1 & 8:14:05.4 & +  70:52:43 & 22.342 $\pm$ 0.288 & -0.135 $\pm$ 0.609 &  0.038 $\pm$ 0.659 \nl 
 & & & & & \nl 
 H37-1 & 8:14:07.2 & +  70:54:57 & 21.519 $\pm$ 0.053 &  0.320 $\pm$ 0.080 &  0.629 $\pm$ 0.079 \nl 
 & & & & & \nl 
 H41-1 & 8:14:12.2 & +  70:50:41 & 22.430 $\pm$ 0.086 &  0.405 $\pm$ 0.131 &  1.133 $\pm$ 0.111 \nl 
 & & & & & \nl 
 H42-1 & 8:14:08.4 & +  70:54:45 & 21.787 $\pm$ 0.079 &  0.364 $\pm$ 0.125 &  0.588 $\pm$ 0.124 \nl 
 & & & & & \nl 
 H44-1 & 8:14:11.7 & +  70:52:60 & 22.039 $\pm$ 0.082 &  0.717 $\pm$ 0.113 &  1.553 $\pm$ 0.099 \nl 
 H44-2 & 8:14:13.0 & +  70:52:51 & 21.053 $\pm$ 0.045 &  0.049 $\pm$ 0.083 &  0.421 $\pm$ 0.076 \nl 
 & & & & & \nl 
 H46-1 & 8:14:15.9 & +  70:50:28 & 22.523 $\pm$ 0.098 &  1.357 $\pm$ 0.111 &  2.119 $\pm$ 0.104 \nl 
 & & & & & \nl 
 H48-1 & 8:14:21.4 & +  70:54:23 & 22.040 $\pm$ 0.198 & -0.682 $\pm$ 0.525 & -0.155 $\pm$ 0.412 \nl 
 H48-2 & 8:14:21.8 & +  70:54:31 & 21.970 $\pm$ 0.095 &  0.275 $\pm$ 0.146 &  0.817 $\pm$ 0.130 \nl 
 H48-3 & 8:14:21.4 & +  70:54:15 & 20.378 $\pm$ 0.043 & -0.276 $\pm$ 0.082 & -0.206 $\pm$ 0.085 \nl 
 & & & & & \nl 
 H51-1 & 8:14:34.0 & +  70:55:18 & 22.375 $\pm$ 0.074 &  1.685 $\pm$ 0.083 &  2.868 $\pm$ 0.077 
\enddata
\end{deluxetable}

\clearpage
\begin{deluxetable}{cl}
\tablecaption{Summary of \hi\ Hole Classification}
\label{table:photsummary}
\tablewidth{400pt}
\tablenum{4}
\tablehead{
\colhead{Hole Type} & \colhead{Hole Numbers}
}
\startdata
1 &	1,\hskip .27cm   2,\hskip .27cm   3,\hskip .27cm   4,\hskip .27cm   7,\hskip .27cm   8,\hskip .27cm   10,\hskip .27cm   11,\hskip .27cm   13,\hskip .27cm   15,\hskip .27cm   18,\hskip .27cm   27,\hskip .27cm   40,\hskip .27cm   47,\hskip .27cm   49,\hskip .27cm   50 \nl

2 &	14,\hskip .27cm   20,\hskip .27cm   23,\hskip .27cm   24,\hskip .27cm   28,\hskip .27cm   30,\hskip .27cm   31,\hskip .27cm   32,\hskip .27cm   33,\hskip .27cm   34,\hskip .27cm   35,\hskip .27cm   39,\hskip .27cm   45 \nl

3 &	16,\hskip .27cm   21,\hskip .27cm   22,\hskip .27cm   36,\hskip .27cm   44,\hskip .27cm   48 \nl

4 &	43 \nl

5 &	9,\hskip .27cm   12,\hskip .27cm   19,\hskip .27cm   37,\hskip .27cm   41,\hskip .27cm   42,\hskip .27cm   46,\hskip .27cm   51 \nl
	
6 &	5,\hskip .27cm   6,\hskip .27cm   17,\hskip .27cm   25,\hskip .27cm   26,\hskip .27cm   29,\hskip .27cm   38 \nl
\enddata
\tablecomments{ Holes are classified into one of six
categories (see Section 2.3 for details):
(1) Empty Hole;
(2) Galaxian Background;
(3) Possible Star Cluster;
(4) Possible Photoionization Region;
(5) Faint Foreground Star;
(6) Contaminated/No Photometry.
}
\end{deluxetable}

\clearpage
\begin{deluxetable}{lrrcrcr}
\tablewidth{360pt}
\tablenum{5}
\tablecaption{Predicted Parameters for Putative Clusters in \hi\ Holes}
\label{table:energy.out}
\tablehead{
\colhead{} & \colhead{} & \colhead{} & 
\multicolumn{2}{c}{Model I \tablenotemark{1}} & 
\multicolumn{2}{c}{Model II \tablenotemark{2}} \\
\cline{4-5} \cline{6-7} \\
\colhead{Hole \#} &      \colhead{Hole Energy} &
\colhead{Hole Age} & \colhead{B} & 
\colhead{B$-$V} & \colhead{B} & 
\colhead{B$-$V} \\
\colhead{} & \colhead{($10^{50}$ ergs)} &
\colhead{($10^6$ yr)} & \colhead{} & 
\colhead{} & \colhead{} & 
\colhead{} 
}
\startdata
  2 &  15.5\ \ \ \ \ \ \ & 101.6 \ \ \ \ & 25.98 &  0.059 \ & 25.16 &  0.081 \ \nl 
  5 & 180.6\ \ \ \ \ \ \ &  61.4 \ \ \ \ & 23.04 &  0.025 \ & 22.27 &  0.048 \ \nl 
  8 & 129.0\ \ \ \ \ \ \ & 128.2 \ \ \ \ & 23.84 &  0.080 \ & 23.00 &  0.101 \ \nl 
  9 &  48.8\ \ \ \ \ \ \ &  86.5 \ \ \ \ & 24.63 &  0.046 \ & 23.82 &  0.068 \ \nl 
 10 & 318.9\ \ \ \ \ \ \ &  58.4 \ \ \ \ & 22.39 &  0.021 \ & 21.62 &  0.044 \ \nl 
 12 &  11.7\ \ \ \ \ \ \ &  19.9 \ \ \ \ & 25.14 & -0.075 \ & 24.54 & -0.047 \ \nl 
 13 & 406.2\ \ \ \ \ \ \ &  97.6 \ \ \ \ & 22.41 &  0.056 \ & 21.59 &  0.078 \ \nl 
 14 &   9.4\ \ \ \ \ \ \ &  47.3 \ \ \ \ & 26.07 &  0.002 \ & 25.33 &  0.027 \ \nl 
 17 &  22.6\ \ \ \ \ \ \ & 127.9 \ \ \ \ & 25.73 &  0.080 \ & 24.89 &  0.101 \ \nl 
 19 &  18.3\ \ \ \ \ \ \ &  90.9 \ \ \ \ & 25.72 &  0.050 \ & 24.91 &  0.072 \ \nl 
 21 & 1747.1\ \ \ \ \ \ \ & 121.1 \ \ \ \ & 20.97 &  0.074 \ & 20.13 &  0.096 \ \nl 
 22 &  30.2\ \ \ \ \ \ \ &  34.3 \ \ \ \ & 24.55 & -0.026 \ & 23.85 & -0.001 \ \nl 
 23 &  35.5\ \ \ \ \ \ \ &  60.4 \ \ \ \ & 24.80 &  0.023 \ & 24.02 &  0.047 \ \nl 
 25 &  32.0\ \ \ \ \ \ \ &  90.9 \ \ \ \ & 25.12 &  0.050 \ & 24.31 &  0.072 \ \nl 
 27 &  57.6\ \ \ \ \ \ \ &  90.1 \ \ \ \ & 24.47 &  0.049 \ & 23.67 &  0.071 \ \nl 
 28 &   6.9\ \ \ \ \ \ \ &  18.4 \ \ \ \ & 25.66 & -0.083 \ & 25.07 & -0.054 \ \nl 
 29 &  13.3\ \ \ \ \ \ \ &  52.7 \ \ \ \ & 25.76 &  0.012 \ & 25.01 &  0.036 \ \nl 
 30 & 460.0\ \ \ \ \ \ \ &  58.7 \ \ \ \ & 22.00 &  0.021 \ & 21.23 &  0.045 \ \nl 
 32 &  30.3\ \ \ \ \ \ \ &  53.3 \ \ \ \ & 24.88 &  0.013 \ & 24.12 &  0.037 \ \nl 
 33 &   6.2\ \ \ \ \ \ \ &  48.6 \ \ \ \ & 26.54 &  0.005 \ & 25.79 &  0.029 \ \nl 
 34 &  16.4\ \ \ \ \ \ \ &  39.0 \ \ \ \ & 25.31 & -0.015 \ & 24.60 &  0.010 \ \nl 
 35 &  64.2\ \ \ \ \ \ \ &  27.0 \ \ \ \ & 23.54 & -0.048 \ & 22.88 & -0.021 \ \nl 
 36 &  74.4\ \ \ \ \ \ \ &  30.6 \ \ \ \ & 23.48 & -0.037 \ & 22.80 & -0.010 \ \nl 
 37 &  17.3\ \ \ \ \ \ \ &  11.5 \ \ \ \ & 24.28 & -0.126 \ & 23.78 & -0.094 \ \nl 
 38 &  46.1\ \ \ \ \ \ \ &  64.0 \ \ \ \ & 24.50 &  0.022 \ & 23.73 &  0.045 \ \nl 
 42 &  20.5\ \ \ \ \ \ \ &  15.4 \ \ \ \ & 24.33 & -0.099 \ & 23.77 & -0.069 \ \nl 
 43 &  30.2\ \ \ \ \ \ \ &  23.3 \ \ \ \ & 24.24 & -0.061 \ & 23.60 & -0.033 \ \nl 
 44 & 135.8\ \ \ \ \ \ \ &  41.8 \ \ \ \ & 23.07 & -0.009 \ & 22.35 &  0.016 \ \nl 
 45 &  33.0\ \ \ \ \ \ \ &  33.0 \ \ \ \ & 24.42 & -0.030 \ & 23.73 & -0.004 \ \nl 
 47 & 646.4\ \ \ \ \ \ \ &  99.6 \ \ \ \ & 21.91 &  0.057 \ & 21.10 &  0.079 \ \nl 
 48 & 400.4\ \ \ \ \ \ \ &  24.5 \ \ \ \ & 21.47 & -0.057 \ & 20.83 & -0.029 \ \nl 
 49 & 357.6\ \ \ \ \ \ \ &  86.2 \ \ \ \ & 22.46 &  0.046 \ & 21.66 &  0.068 \ \nl 
 50 & 352.2\ \ \ \ \ \ \ &  23.9 \ \ \ \ & 21.59 & -0.059 \ & 20.95 & -0.031 \ \nl 
 51 & 355.7\ \ \ \ \ \ \ & 100.5 \ \ \ \ & 22.57 &  0.058 \ & 21.75 &  0.080 \ \nl 
\enddata
\tablenotetext{1}{ Salpeter IMF; stars with masses $\geq$7 $M_{\sun}$
become supernovae}
\tablenotetext{2}{ Miller-Scalo IMF; stars with masses $\geq$8 $M_{\sun}$
become supernovae}
\end{deluxetable}

\clearpage
\begin{deluxetable}{ccrrcrcrccr}
\tablewidth{490pt}
\tablenum{6}
\tablecaption{Comparison of Model Clusters with Observed Quantities}
\label{table:comparison}
\tablehead{
\colhead{} & \colhead{} & \colhead{} & \colhead{} &
\multicolumn{4}{c}{PREDICTED} & 
\multicolumn{3}{c}{OBSERVED} \\
\cline{5-8} \cline{9-11} \\
\colhead{} & \colhead{} & \colhead{} & \colhead{} & 
\multicolumn{2}{c}{Model I \tablenotemark{1}} & 
\multicolumn{2}{c}{Model II \tablenotemark{2}} & \colhead{} & \colhead{}
& \colhead{} \\
\cline{5-6} \cline{7-8} \\
\colhead{Hole \#} &   \colhead{Hole} &   \colhead{Energy} &
\colhead{Age} & \colhead{B} & 
\colhead{B$-$V} & \colhead{B} & 
\colhead{B$-$V} & \colhead{$\mu_B$} & \colhead{B} & \colhead{B$-$V}\\
\colhead{} & \colhead{Type \tablenotemark{3}} & \colhead{($10^{50}$ ergs)} &
\colhead{($10^6$ yr)} & \colhead{} & 
\colhead{} & \colhead{} & 
\colhead{} & \colhead{} & \colhead{} & \colhead{}
}
\startdata
 \ 8 & 1 & 129.0\ \ \ \ \ \ \ & 128.2 \ \ \ \ & 23.84 &  0.080 \ & 23.00 &  0.101 \ & 27.0 & $>$23.0 &  ...... \nl 
 10 & 1 & 318.9\ \ \ \ \ \ \ &  58.4 \ \ \ \ & 22.39 &  0.021 \ & 21.62 &  0.044 \ & 26.3 & $>$23.0 & ...... \nl 
 13 & 1 & 406.2\ \ \ \ \ \ \ &  97.6 \ \ \ \ & 22.41 &  0.056 \ & 21.59 &  0.078 \ & 27.5 & $>$23.0 & ......\nl 
 47 & 1 & 646.4\ \ \ \ \ \ \ &  99.6 \ \ \ \ & 21.91 &  0.057 \ & 21.10 &  0.079 \ & 27.6 & $>$23.0 & ...... \nl 
 49 & 1 & 357.6\ \ \ \ \ \ \ &  86.2 \ \ \ \ & 22.46 &  0.046 \ & 21.66 &  0.068 \ & 26.8 & $>$23.0 & ...... \nl
 50 & 1 & 352.2\ \ \ \ \ \ \ &  23.9 \ \ \ \ & 21.59 & -0.059 \ & 20.95 & -0.031 \ & 26.7 & $>$23.0 & ...... \nl\nl
 30 & 2 & 460.0\ \ \ \ \ \ \ &  58.7 \ \ \ \ & 22.00 &  0.021 \ & 21.23 &  0.045 \ & 24.1 & 21.75 & 1.171 \nl 
 35 & 2 & 64.2\ \ \ \ \ \ \ &  27.0 \ \ \ \ & 23.54 & -0.048 \ & 22.88 & -0.021 \ & 25.6 & 20.39 & 1.594 \nl \nl
 21 & 3 & 1747.1\ \ \ \ \ \ \ & 121.1 \ \ \ \ & 20.97 &  0.074 \ & 20.13 &  0.096 \ & 23.9 & 20.90 & -0.089 \nl
 36 & 3 & 74.4\ \ \ \ \ \ \ &  30.6 \ \ \ \ & 23.48 & -0.037 \ & 22.80 & -0.010 \ & 23.6 & 22.34 & -0.135 \nl 
 44 & 3 & 135.8\ \ \ \ \ \ \ &  41.8 \ \ \ \ & 23.07 & -0.009 \ & 22.35 &  0.016 \ & 24.1 & 21.05 & 0.049  \nl 
 48 & 3 & 400.4\ \ \ \ \ \ \ &  24.5 \ \ \ \ & 21.47 & -0.057 \ & 20.83 & -0.029 \ & 24.8 & 20.38 & -0.276 \nl \nl
 51 & 5 & 355.7\ \ \ \ \ \ \ & 100.5 \ \ \ \ & 22.57 &  0.058 \ & 21.75 &  0.080 \ & 26.7 & 22.38 & 1.685 \nl\nl
 \ 5 & 6 & 180.6\ \ \ \ \ \ \ &  61.4 \ \ \ \ & 23.04 &  0.025 \ & 22.27 &  0.048 \ & ...... & ...... & ...... \nl 
\enddata
\tablenotetext{1}{ Salpeter IMF; stars with masses $\geq$7 $M_{\sun}$
become supernovae}
\tablenotetext{2}{ Miller-Scalo IMF; stars with masses $\geq$8 $M_{\sun}$
become supernovae}
\tablenotetext{3} {Holes are classified into one of six
categories (see Section 2.3 for details):
(1) Empty Hole;
(2) Galaxian Background;
(3) Possible Star Cluster;
(4) Possible Photoionization Region;
(5) Faint Foreground Star;
(6) Contaminated/No Photometry.}
\end{deluxetable}

\clearpage

\figcaption[Rhode.fig1.ps]{\hi\ map of
Ho$\thinspace$II, derived from the data set of P92.  The locations
of the \hi\ holes are indicated, using the numbering scheme of P92.
The size of each pair of concentric circles indicates the sizes of the
apertures and annuli used in our photometric measurements.  Only those
holes for which photometric measurements were carried out are shown.
\label{fig:hi map}}

\figcaption[Rhode.fig2.ps]{Composite
\bvr\ image made by merging $B$, $V$, and $R$-band data from April
1995, shown on the same scale as the \hi\ map in Figure \ref{fig:hi
map}.  The locations of the \hi\ holes are indicated by concentric
circles, which are identical to those in Figure 1.
\label{fig:combined BVR image}}

\figcaption[Rhode.fig3.ps]
{Continuum-subtracted H$\alpha$ image, shown on the same scale as the
images in Figures \ref{fig:hi map} and \ref{fig:combined BVR image},
with the locations of the \hi \ holes marked.  Note that, with the
exception of holes \#16, 20, and 43, none of the \hi\ holes are
coincident with \hii\ regions.  In addition, none of the holes that
are not associated with \hii\ regions contain any detectable diffuse
H$\alpha$ emission.  Some patches of diffuse emission are seen {\it
between} some of the holes.
\label{fig:HA image}}

\end{document}